\documentstyle[amssymb,aps,12pt]{revtex}

\begin{document}
\author{Rong Cheng and J.-Q. Liang}
\title{Superfluidity of spin-1 bosons in optical lattices}
\address{Institute of Theoretical Physics, Shanxi University, Taiyuan 030006, China}
\maketitle

\begin{abstract}
In this paper we show that the superfluidity of cold spin--1 Bose atoms of
weak interactions in an optical lattice can be realized according to the
excitation energy spectrum which is derived by means of Bogliubov
transformation. The characteristic of the superfluid-phase spectrum is
explained explicitly in terms of the nonvanishing critical velocity, i.e.,
the Landau criterion. It is observed that critical velocities of superfluid
are different for three spin components and, moreover, can be controlled by
adjusting the lattice parameters in practical experiments to detect the
superfluid phase.
\end{abstract}


\section{INTRODUCTION}

Originally discovered in the system of liquid helium and later in the
context of superconductors, superfluidity is a hallmark property of
interacting quantum fluids and encompasses a whole class of fundamental
phenomena such as the absence of viscosity, persistent currents and
quantized vortices. With the achievement of Bose-Einstein condensation (BEC)
in alkali-metal atoms, the weakly interacting Bose gases have served as an
idealized model of the superfluid \cite{1} and a test ground of macroscopic
quantum effects at low temperatures. Recently, rapid advances of
experimental techniques in optical traps \cite{2,3,4} open up a prospect for
the study of the superfluidity of BEC trapped in periodic potentials, which
has attracted fast growing interests both experimentally and theoretically 
\cite{5,6}. The main reason is that the optical lattices possess
controllable potential depths and lattice constants by adjusting the
intensity of the laser beams in realistic experiments and moreover the
lattice is basically defect free. In addition, the great advantage of
optical traps is that it liberates the spin degree of freedom and provides
us an opportunity to test spin-dependent quantum phenomena that are absent
in the scale-condensate cases. Theoretical studies have predicted a variety
of novel phenomena of spinor condensates such as fragmented condensation 
\cite{7}, skyrmion excitations \cite{8} and propagation of spin waves \cite
{8,9}. The quantum phase transition from a superfluid to a Mott-insulator
(SF-MI) phase in spinor BEC has been observed in experiments \cite{4}.{\sl \ 
}In the Mott-insulator phase (MIP), atoms are localized; the particle-number
fluctuations at each lattice site are suppressed so that there is no phase
coherence across the lattice. When the tunnel coupling through the interwell
barriers becomes large compared to the atom--atom interactions, the system
undergoes a phase transition into the superfluid phase (SFP) in which the
atom number per site is random and hence wave function exhibits long-range
phase coherence. That means one can go from the regime in which the
interaction energy dominates (high barrier of periodic potential) to the
regime where the kinetic energy is the leading part (low barrier of periodic
potential) by varying the intensity of laser beams and vice versa. The SF-MI
transition has been{\sl \ }also investigated in Refs.\cite{10,11,12,13,14},
where the Bose--Hubbard model is introduced as the starting point, and the
analytic phase-transition condition and phase diagram have been obtained by
using a perturbation expansion over the superfluid order parameter with
on-site zero-order energy spectrum which, although gives rise to a
reasonable description of the MIP, the SFP is not described explicitly{\sl 
{\bf .} }In the present paper we study the spinor BEC in a parameter region
such that its ground state stands deeply in the SFP \cite{14}. The Bogliubov
approach is used to obtain the explicit excitation energy spectra of weakly
interacting spin--1 atoms in an optical lattice and the superfluidity is
explained explicitly in terms of the energy spectra. The critical velocities
known as the Landau criterion for the superfluid phase are evaluated for the 
$^{23}$Na atoms and are seen to be realizable in practical experiments. It
is also demonstrated that the critical velocities are{\bf \ }spin component
dependent and controllable by adjusting of the lattice parameters{\bf .} Our
result may throw light on the experimental observation of the persistent
atom-current density for the spinor-atom matter waves.

\section{BOGLIUBOV METHOD AND ENERGY SPECTRUM}

Alkali-metal atoms with nuclear spin $I=3/2,$ such as $^{23}$Na, $^{39}$K,
and $^{87}$Rb, behave at low temperatures like simple bosons with a
hyperfine spin $f=1$. The most general model Hamiltonian for the dilute gas
of bosonic atoms with hyperfine spin $f=1$ trapped in the optical potential
can be written, in the second-quantization notation, as 
\begin{eqnarray}
\widehat{H} &=&\sum_\alpha \int d^3X\widehat{\psi }_\alpha ^{\dagger }\left(
X\right) \left( -\frac{\nabla ^2}{2M}+V_0\left( X\right) +V_T\left( X\right)
\right) \widehat{\psi }_\alpha \left( X\right)  \nonumber \\
&&+\frac{C_0}2\sum_{\alpha ,\beta }\int d^3X\widehat{\psi }_\alpha ^{\dagger
}\left( X\right) \widehat{\psi }_\beta ^{\dagger }\left( X\right) \widehat{%
\psi }_\beta \left( X\right) \widehat{\psi }_\alpha \left( X\right)
\label{1} \\
&&+\frac{C_2}2\sum_{\alpha ,\beta ,\alpha ^{^{\prime }},\beta ^{^{\prime
}}}\int d^3X\widehat{\psi }_\alpha ^{\dagger }\left( X\right) \widehat{\psi }%
_\beta ^{\dagger }\left( X\right) F_{\alpha \alpha ^{^{\prime }}}F_{\beta
\beta ^{^{\prime }}}\widehat{\psi }_{\beta ^{^{\prime }}}\left( X\right) 
\widehat{\psi }_{\alpha ^{^{\prime }}}\left( X\right) ,  \nonumber
\end{eqnarray}
where $M$ is the mass of a single atom; $\widehat{\psi }_\alpha \left(
X\right) $ is the atomic field annihilation operator associated with atoms
in the hyperfine spin state $|f=1,m_f=\alpha \rangle $ and the indices $%
\alpha ,\beta ,\alpha ^{^{\prime }},\beta ^{^{\prime }}$ label the three
spin components $\left( \alpha ,\beta ,\alpha ^{^{\prime }},\beta ^{^{\prime
}}=-1,0,1\right) $. $V_0\left( X\right) =V_0\left( \sin ^2kX_1+\sin
^2kX_2+\sin ^2kX_3\right) $ is the optical lattice potential formed by laser
beams, which is assumed to be the same for all three spin components, where $%
k=2\pi /\lambda $ is the wave vector of the laser light with $\lambda $
being the wavelength of the laser light and $V_0$ is the tunable depth of
the potential well and, hence, the lattice constant is $d=\lambda /2$. $%
V_T\left( X\right) $ denotes an additional (slowly varying) external
trapping potential, e.g., a magnetic trap. The $3\times 3$ spin matrices $%
{\sl F}$\ denote the conventional three-dimensional representation
(corresponding to the spin ${\sl f=1}$) of the angular momentum operator
with ${\sl F}_{\alpha \beta }^x=\left( \delta _{\alpha ,\beta -1}+\delta
_{\alpha ,\beta +1}\right) /\sqrt{2},$ ${\sl F}_{\alpha \beta }^y=i\left(
\delta _{\alpha ,\beta -1}-\delta _{\alpha ,\beta +1}\right) /\sqrt{2},$ $%
{\sl F}_{\alpha \beta }^z=\alpha \delta _{\alpha \beta }.$ The coefficients $%
C_0$ and $C_2$ are related to scattering lengths $a_0$ and $a_2$ of two
colliding bosons with total angular momenta $0$ and $2$, respectively, by $%
C_0=4\pi \hbar ^2\left( 2a_2+a_0\right) /3M$ and $C_2=4\pi \hbar ^2\left(
a_2-a_0\right) /3M$. For atoms $^{23}$Na, we have $a_2>a_0$, that is $C_2>0$%
\ and the interaction is antiferromagnetic.{\rm \ }While, for $^{87}$Rb
atoms the situation is just opposite that $a_2<a_0$\ (this leads to $C_2<0$%
)\ and the interaction is ferromagnetic \cite{8}. For the periodic
potential, the energy eigenstates are Bloch states. We can expand the field
operator $\widehat{\psi }_\alpha \left( X\right) $ in the Wannier basis,
which is a superposition of Bloch states such that 
\begin{equation}
\widehat{\psi }_\alpha \left( X\right) =\sum_i\widehat{a}_{\alpha i}w\left(
X-X_i\right) ,  \label{2}
\end{equation}
where $w\left( X-X_i\right) $ are the Wannier functions localized in the
lattice site $i$ and $\widehat{a}_{\alpha i}$ corresponds to the bosonic
annihilation operator on the $i$th lattice site. Also Eq. (\ref{2}) suggests
that atoms in different spin states are approximately described by the same
coordinate wave function, which is seen to be the case when the
spin-symmetric interaction is strong compared with the asymmetric part,
i.e., $\left| C_0\right| \gg \left| C_2\right| $ \cite{11}. This is relevant
to experimental conditions for $^{23}$Na and $^{87}$Rb atoms. Using Eq. (\ref
{2}), the general Hamiltonian (1) reduces to the Bose--Hubbard Hamiltonian 
\begin{eqnarray}
\widehat{H} &=&-J\sum_{\left\langle i,j\right\rangle }\sum_\alpha \widehat{a}%
_{\alpha i}^{\dagger }\widehat{a}_{\alpha j}+\sum_i\sum_\alpha \varepsilon _i%
\widehat{a}_{\alpha i}^{\dagger }\widehat{a}_{\alpha i}  \nonumber \\
&&+\frac 12U_0\sum_i\sum_{\alpha ,\beta }\widehat{a}_{\alpha i}^{\dagger }%
\widehat{a}_{\beta i}^{\dagger }\widehat{a}_{\beta i}\widehat{a}_{\alpha i}
\label{3} \\
&&+\frac 12U_2\sum_i\sum_{\alpha ,\beta ,\alpha ^{^{\prime }},\beta
^{^{\prime }}}\widehat{a}_{\alpha i}^{\dagger }\widehat{a}_{\beta
i}^{\dagger }F_{\alpha \alpha ^{^{\prime }}}F_{\beta \beta ^{^{\prime }}}%
\widehat{a}_{\beta ^{^{\prime }}i}\widehat{a}_{\alpha ^{^{\prime }}i}. 
\nonumber
\end{eqnarray}
Here the first term in Eq. (\ref{3}) describes the strength of the
spin-symmetric tunneling, which is characterized by the hopping matrix
element between adjacent sites$\ i$\ and $j.$\ The tunneling constant $%
J=-\int d^3Xw^{*}\left( X-X_i\right) \left[ -\nabla ^2/2M+V_0\left( X\right)
\right] w\left( X-X_j\right) $ depends exponentially on the depth of
potential well $V_0$\ and can be varied experimentally by several orders of
magnitude. The second term denotes energy offset of the$\ i$th lattice site
due to the external confinement of the atoms, where the parameter $%
\varepsilon _i=\int d^3Xw^{*}\left( X-X_i\right) V_T\left( X\right) w\left(
X-X_i\right) $\ is assumed to be of the same value $\varepsilon $\ for all
lattice sites in the present paper. The third and fourth terms characterize
the repulsion interaction between two atoms on a single lattice site, which
is quantified by the on-site interaction matrix element $U_{0(2)}=C_{0(2)}%
\int \left| w\left( X-X_i\right) \right| ^4d^3X$. In our case the
interaction energy is very well determined by the single parameters $U_0$
and $U_2$, due to the short range of the interactions, which is smaller than
the lattice spacing.

Now we use Bogliubov approach to diagonalize the Hamiltonian and obtain the
excitation energy spectrum of spin--1 bosons with weak{\rm \ }interaction 
\cite{14} in an optical lattice and hence to study the SFP property
explicitly. To this end we firstly express the site space operator $\widehat{%
a}_{\alpha i}$ in terms of the wave--vector space operator $\widehat{a}%
_{k,\alpha }$ as 
\begin{equation}
\left\{ 
\begin{array}{c}
\widehat{a}_{\alpha i}=\frac 1{\sqrt{N_s}}\sum_k\widehat{a}_{k,\alpha
}e^{ik\cdot X_i}, \\ 
\widehat{a}_{\alpha i}^{\dagger }=\frac 1{\sqrt{N_s}}\sum_k\widehat{a}%
_{k,\alpha }^{\dagger }e^{-ik\cdot X_i},
\end{array}
\right.   \label{4}
\end{equation}
where $N_s$\ is the total number of the lattice sites and $X_i$\ is the
coordinate of site $i$. The wave vector ${\sl k}$\ runs only over the first
Brillouin zone. In the tight--binding approximation (TBA) if we limit our
description to simple cubic lattice and substitute Eq. (\ref{4}) into the
Hamiltonian (3), we can obtain 
\begin{eqnarray}
\widehat{H} &=&\sum_\alpha \sum_k\varepsilon \left( k\right) \widehat{a}%
_{k,\alpha }^{\dagger }\widehat{a}_{k,\alpha }+\widehat{H}_{int},  \nonumber
\\
\widehat{H}_{int} &=&\frac{U_0}{2N_s}\sum_{\alpha ,\beta
}\sum_{k,p,k^{^{\prime }},p^{^{\prime }}}\delta _{k+p,k^{^{\prime
}}+p^{^{\prime }}}\widehat{a}_{k,\alpha }^{\dagger }\widehat{a}_{p,\beta
}^{\dagger }\widehat{a}_{p^{^{\prime }},\beta }\widehat{a}_{k^{^{\prime
}},\alpha }  \label{5} \\
&&+\frac{U_2}{2N_s}\sum_{\alpha ,\beta ,\alpha ^{^{\prime }},\beta
^{^{\prime }}}\sum_{k,p,k^{^{\prime }},p^{^{\prime }}}\delta
_{k+p,k^{^{\prime }}+p^{^{\prime }}}\widehat{a}_{k,\alpha }^{\dagger }%
\widehat{a}_{p,\beta }^{\dagger }F_{\alpha \alpha ^{^{\prime }}}F_{\beta
\beta ^{^{\prime }}}\widehat{a}_{p^{^{\prime }},\beta ^{^{\prime }}}\widehat{%
a}_{k^{^{\prime }},\alpha ^{^{\prime }}},  \nonumber
\end{eqnarray}
where $\varepsilon \left( k\right) =\varepsilon -Jz\cos \left( kd\right) $\
with $z$\ the number of nearest neighbors of each site. Since the number of
atoms condensed in the zero-momentum state is much larger than one, we have 
\begin{equation}
\sum_\alpha \widehat{a}_{\alpha 0}\widehat{a}_{\alpha 0}^{\dagger
}=\sum_\alpha \widehat{a}_{\alpha 0}^{\dagger }\widehat{a}_{\alpha
0}+1\approx \sum_\alpha N_{\alpha 0}={\cal N}_0\gg 1.  \label{6}
\end{equation}
$N_{\alpha 0}$ is the number of condensed atoms of spin-$\alpha $ component
in the zero-momentum state and ${\cal N}_0$ is the total number of condensed
atoms. So, we can replace the operator $\widehat{a}_{\alpha 0}$ and $%
\widehat{a}_{\alpha 0}^{\dagger }$ with a $``$ $c$ $"$ number $\sqrt{%
N_{\alpha 0}}$. Thus we have 
\begin{equation}
N_{\alpha 0}=N_\alpha -\sum_{k\neq 0}\widehat{a}_{k,\alpha }^{\dagger }%
\widehat{a}_{k,\alpha },  \label{7}
\end{equation}
where $N_\alpha $ is the total number of atoms of spin-$\alpha $ component
and the term of $k\neq 0$ is exclusive in the wave-vector sum. Moreover, in
the interacting part of the Hamiltonian when $k\neq 0$ we regard $\widehat{a}%
_{k,\alpha }^{\dagger }$, $\widehat{a}_{k,\alpha }$ as the small deviation
from the operators of vanishing momentum, thus all products of four boson
operators are approximated as quadratic form, for example, 
\begin{equation}
\widehat{a}_{\ 0,\alpha }^{\dagger }\widehat{a}_{0,\alpha }^{\dagger }%
\widehat{a}_{0,\alpha }\widehat{a}_{0,\alpha }=N_{\alpha 0}^2\approx
N_\alpha ^2-2N_\alpha \sum_{k\neq 0}\widehat{a}_{k,\alpha }^{\dagger }%
\widehat{a}_{k,\alpha },  \label{8}
\end{equation}
\begin{equation}
\sum_{\alpha ,\beta ,k\neq 0}\widehat{a}_{\ 0,\alpha }^{\dagger }\widehat{a}%
_{k,\beta }^{\dagger }\widehat{a}_{k,\beta }\widehat{a}_{0,\alpha
}=\sum_{\alpha ,\beta ,k\neq 0}\widehat{a}_{\ k,\beta }^{\dagger }\widehat{a}%
_{k,\beta }N_{\alpha 0}=\left[ \sum_\alpha \left( N_\alpha -\sum_{k\neq 0}%
\widehat{a}_{k,\alpha }^{\dagger }\widehat{a}_{k,\alpha }\right) \right]
\sum_{\beta ,k\neq 0}\widehat{a}_{\ k,\beta }^{\dagger }\widehat{a}_{k,\beta
}\approx N\sum_{\alpha ,k\neq 0}\widehat{a}_{\ k,\alpha }^{\dagger }\widehat{%
a}_{k,\alpha },  \label{9}
\end{equation}
where $N=\sum_\alpha N_\alpha $ is the total number of atoms. With the
approximation Eqs. (\ref{8}) and (\ref{9}) the total Hamiltonian (5) can be
written as 
\begin{eqnarray}
\widehat{H} &=&\frac{U_0}{2N_s}N^2+N\left( \varepsilon -zJ\right) +\frac{U_2%
}{2N_s}[\left( N_1-N_{-1}\right) ^2+2N_0(\sqrt{N_1}+\sqrt{N_{-1}}%
)^2]+\sum_{k\neq 0}\left[ \sum_\alpha \varepsilon \left( k\right) \widehat{a}%
_{\ k,\alpha }^{\dagger }\widehat{a}_{k,\alpha }\right.   \nonumber \\
&&+\frac{U_0}{2N_s}\sum_{\alpha ,\beta }\sqrt{N_{\alpha 0}}\sqrt{N_{\beta 0}}%
(\widehat{a}_{\ k,\alpha }^{\dagger }\widehat{a}_{-k,\beta }^{\dagger }+%
\widehat{a}_{k,\alpha }\widehat{a}_{-k,\beta }+2\widehat{a}_{\ k,\alpha
}^{\dagger }\widehat{a}_{k,\beta })+\frac{U_2}{2N_s}\left( 2N_0(\widehat{a}%
_{k,1}^{\dagger }\widehat{a}_{-k,-1}^{\dagger }+\widehat{a}_{-k,1}\widehat{a}%
_{k,-1})\right. +\sum_{\gamma =\pm 1}N_\gamma (\widehat{a}_{k,\gamma
}^{\dagger }\widehat{a}_{-k,\gamma }^{\dagger }  \nonumber \\
&&+\widehat{a}_{k,\gamma }\widehat{a}_{-k,\gamma }+2\widehat{a}_{k,\gamma
}^{\dagger }\widehat{a}_{k,\gamma })+2\sum_{\gamma =\pm 1}\sqrt{N_{\gamma 0}}%
\sqrt{N_{00}}(\widehat{a}_{k,\gamma }^{\dagger }\widehat{a}_{-k,0}^{\dagger
}+\widehat{a}_{-k,\gamma }\widehat{a}_{k,0}+\widehat{a}_{k,\gamma }^{\dagger
}\widehat{a}_{k,0}+\widehat{a}_{k,0}^{\dagger }\widehat{a}_{k,\gamma }+2%
\widehat{a}_{k,0}^{\dagger }\widehat{a}_{k,-\gamma }+2\widehat{a}_{k,-\gamma
}^{\dagger }\widehat{a}_{k,0})  \nonumber \\
&&\left. \left. +2\sqrt{N_{10}}\sqrt{N_{-10}}(\widehat{a}_{k,0}^{\dagger }%
\widehat{a}_{-k,0}^{\dagger }-\widehat{a}_{k,1}^{\dagger }\widehat{a}%
_{-k,-1}^{\dagger }+\widehat{a}_{k,0}\widehat{a}_{-k,0}-\widehat{a}_{k,-1}%
\widehat{a}_{-k,1}-\widehat{a}_{k,1}^{\dagger }\widehat{a}_{k,-1}-\widehat{a}%
_{k,-1}^{\dagger }\widehat{a}_{k,1}-2\widehat{a}_{k,0}^{\dagger }\widehat{a}%
_{k,0})\right) \right] .  \label{10}
\end{eqnarray}
Note that the ratio $U_2/U_0$\ is proportional to the ratio of $C_2/C_0$\
for all lattice geometries and hence $U_2/U_0$\ is small enough in general.
Therefore the spin-asymmetric part of the interaction is much smaller than
the spin-symmetric one.

The Hamiltonian (10) is quadratic in the operators $\widehat{a}_{k,a}$, $%
\widehat{a}_{-k,a}^{\dagger }$ and can be diagonalized by the linear
transformation 
\begin{eqnarray}
\widehat{a}_{k,a} &=&u_{k,\alpha }\widehat{b}_{k,\alpha }-\upsilon
_{k,\alpha }\widehat{b}_{-k,\alpha }^{\dagger },  \nonumber \\
\widehat{a}_{k,a}^{+} &=&u_{k,\alpha }\widehat{b}_{k,\alpha }^{\dagger
}-\upsilon _{k,\alpha }\widehat{b}_{-k,\alpha },  \label{11}
\end{eqnarray}
known as the Bogoliubov transformation. This transformation introduces a new
set of operators $\widehat{b}_{k,\alpha }$ and $\widehat{b}_{k,\alpha
}^{\dagger }$ to which we impose the same Bose-operator commutation
relations $\left[ \widehat{b}_{k,\alpha },\widehat{b}_{k^{^{\prime }},\beta
}^{\dagger }\right] =\delta _{k,k^{^{\prime }}}\delta _{\alpha \beta }$ . It
is easy to check that the commutation relations are fulfilled if the
parameters $u_{k,\alpha }$ and $\upsilon _{k,\alpha }$ satisfy the relation 
\begin{equation}
u_{k,\alpha }^2-\upsilon _{k,\alpha }^2=1,  \label{12}
\end{equation}
where the auxiliary parameters $u_{k,\alpha }$ and $\upsilon _{k,\alpha }$
are to be chosen in order to have the vanishing coefficients of the
nondiagonal terms $\widehat{b}_{k,\alpha }^{\dagger }\widehat{b}_{-k,\alpha
}^{\dagger }$ and $\widehat{b}_{k,\alpha }\widehat{b}_{-k,\alpha }$ in the
Hamiltonian (10) (see the Appendix for detail). In virtue of the Bogliubov
transformation (11), we finally obtain the diagonalized Hamiltonian as 
\begin{equation}
\widehat{H}=E_c+\sum_{k\neq 0}E_{k,\alpha ,\alpha }\widehat{b}_{k,\alpha
}^{\dagger }\widehat{b}_{k,\alpha }  \label{13}
\end{equation}
with 
\[
E_c=\frac 12U_0Nn^2+N\left( \varepsilon -zJ\right) +\frac{U_2}{2N_s}[\left(
N_1-N_{-1}\right) ^2+2N_0(\sqrt{N_1}+\sqrt{N_{-1}})^2],
\]
where the energy spectra $E_{k,\alpha ,\alpha }$ $\left( \alpha =0,\pm
1\right) $ of the quasiparticle are given by 
\begin{equation}
E_{k,\gamma ,\gamma }=\sqrt{\overline{\varepsilon }_k\left( \overline{%
\varepsilon }_k+2U_0n_\gamma +2U_2n_\gamma \right) },  \label{14}
\end{equation}
\begin{equation}
E_{k,0,0}=\sqrt{\overline{\varepsilon }_k\left( \overline{\varepsilon }%
_k+2U_0n_0+4U_2\sqrt{n_{-1}}\sqrt{n_1}\right) },  \label{15}
\end{equation}
where $\gamma =\pm 1$ and 
\[
\overline{\varepsilon }_{k\left( k\neq 0\right) }=zJ\left[ 1-\cos \left(
kd\right) \right] .
\]
The symbol $n_\alpha =N_\alpha /N_s$ represents the average atom number of
spin-$\alpha $ component per lattice site. In the experiments of Ref. \cite
{4} the number of atoms per lattice site is shown to be around $1-3$. 

\section{CRITICAL\ VELOCITY\ OF\ SUPERFLUID}

The energy spectra Eqs. (\ref{14}) and (\ref{15}) are typical for the
superfluid. To this end we look at the dispersion relations of energy
spectra $E_{k,\alpha ,\alpha }$ for the limit case $k\rightarrow 0$ 
\begin{equation}
E_{k,\gamma ,\gamma }\thicksim \left[ zJd^2\left( U_0+U_2\right) n_\gamma
\right] ^{1/2}k,  \label{16}
\end{equation}
\begin{equation}
E_{k,0,0}\thicksim \left[ zJd^2\left( U_0n_0+2U_2\sqrt{n_{-1}}\sqrt{n_1}%
\right) \right] ^{1/2}k.  \label{17}
\end{equation}
The linear wave--vector dependence of the excitation spectra $E_{k,\alpha
,\alpha }$ is the characteristic of the superfluid which gives rise to
critical velocities of superfluid found as 
\begin{equation}
\upsilon _{s,\gamma }=\left( \frac{\partial E_{k,\gamma ,\gamma }}{\partial k%
}\right) _{k\rightarrow 0}=\frac 1\hbar \left[ zJd^2\left( U_0+U_2\right)
n_\gamma \right] ^{1/2},  \label{18}
\end{equation}
\begin{equation}
\upsilon _{s,0}=\left( \frac{\partial E_{k,0,0}}{\partial k}\right)
_{k\rightarrow 0}=\frac 1\hbar \left[ zJd^2\left( U_0n_0+2U_2\sqrt{n_{-1}}%
\sqrt{n_1}\right) \right] ^{1/2},  \label{19}
\end{equation}
which reduce to the critical velocity of superfluid given in Ref. \cite{13}
for the spin-zero case when{\sl \ }$U_2=0$, where $1/\hbar $ is dimension
correction. The nonvanishing velocity is nothing but the Landau criterion
for the superfluid phase. As seen from the above formulas (18) and (19),
whether there exist critical velocities of superfluid or not{\sl \ }depend
on appropriate values of $J$ and $U_{0(2)}$ which are related to the Wannier
functions determined essentially by the potential of optical lattice. Thus, $%
J$ and $U_{0(2)}$ can be controlled dependently by adjusting the laser
parameters{\sl . }Since the spin-asymmetric interaction $U_2${\rm \ }is
typically one to two orders of magnitude less than the spin-symmetric
interaction $U_0$,{\rm \ }it\ can ensure that the nonvanishing $\upsilon
_{s,\alpha }$\ ($\alpha =0,\pm 1$) exist, whether for the antiferromagnetic
interaction $\left( U_2>0\right) $\ or for the case of ferromagnetic
interaction $\left( U_2<0\right) $.{\bf \ }

Certainly, it is important to see whether or not the superfluid phase can be
realized practically with the recent progress of experiments on the
confinement of atoms in the light-induced trap.{\rm \ }To see this we
evaluate the values of critical velocities of superfluid adopting the
typical experimental data in Ref. \cite{15} for a spin-1 condensate of{\sl \ 
}$^{23}$Na atoms \ in the optical lattice created by three perpendicular
standing laser beams with $\lambda =985$ nm.{\sl \ }The scattering lengths
for $^{23}$Na atoms are ${\sl a}_0{\sl =(}46{\sl \pm }5{\sl )a}_B$ and ${\sl %
a}_2{\sl =(}52\pm 5{\sl )a}_B$, where ${\sl a}_B$ is the Bohr radius
(corresponding to a ratio value{\bf \ }$U_2/U_0=0.04$). The valid condition
of the Bogliubov approach that ${\sl U<<J}$\ (for example, $U/J<<0.1$) can
be fulfilled in a region\ of barrier-height values of the optical lattice
potential $V_0$\ from $0$\ to two or three times $E_R,$\ where $E_R$\ is the
recoil energy{\bf .\ }It turns out that the magnitude of the critical
velocities of the superfluid is the order of mm/s which is seen to be in the
range of experimental values \cite{15}.{\rm \ }

The critical velocities of superfluid $\upsilon _{s,\alpha }$ different for
three spin components are functions of densities. Therefore, the critical
velocities of superfluid can be detected experimentally by counting the
atom--number populations.{\bf \ }Moreover, the component-dependent
velocities may imply component separation of spinor BEC in an optical
lattice similar to the experimentally observed component separation in a
binary mixture of BECs \cite{16}{\bf \ }since one can control the subsequent
time evolution of the mixture of a three-condensate system and detect the
relative motions of the three components which tend to preserve the density
profiles, respectively{\sl . }In particular, we may consider a condensate of
spin--polarized atoms which are all in the state of spin component $\alpha =0
$\ at initial time $t=0$, i.e., $|\psi \left( 0\right) \rangle
=|0,N_0,0\rangle $.{\rm \ }In this case a pair of atoms in the $\alpha =0$
state can be excited into the $\alpha =\pm 1$ states respectively. Thus
after a time $t_c${\rm \ }the number of atoms of the $\alpha =0$\ component
becomes steady such that $N_0\left( t_c\right) =N_0/2$\ \cite{11} and the
number of atoms of $\alpha =\pm 1$\ components is about half of $N_0\left(
t_c\right) $, i.e.,{\rm \ }$N_{-1}\left( t_c\right) =N_1\left( t_c\right)
=N_0\left( t_c\right) /2\approx N_0/4$.{\rm \ }Consequently, the critical
velocities of superfluid for $\alpha =-1$\ $(\upsilon _{s,-1})$\ and $\alpha
=1$\ $\left( \upsilon _{s,1}\right) $\ are equal and different from that for
the $\alpha =0$\ component $\left( \upsilon _{s,0}\right) $.{\rm \ }There
exists a simple relation that $\upsilon _{s,-1}=\upsilon _{s,1}=\upsilon
_{s,0}/\sqrt{2}$\ for the case considered and therefore the atoms of the $%
\alpha =0$\ component can be separated from the mixture of BECs{\bf .}

\section{CONCLUSION}

The energy--band structure of excitation spectra derived in terms of
Bogliubov transformation for spin--1 cold bosons in an optical lattice is
shown to be typical for the superfluid phase from the viewpoint of the
Landau criterion.{\bf \ }Our observation is that the critical velocities of
the superfluid flow are spin-component dependent and can be controlled by
adjusting the laser lights that form the optical lattice. The theoretical
values of critical velocities obtained are in agreement with experimental
observations and possible experiments to detect the superfluid phase are
also discussed{\bf .}

\section{Acknowledgment}

This work was supported by National Natural Science Foundation of China
under Grant Nos. 10475053.

\section{APPENDIX}

The inverse transformation of Eq. (\ref{11}) is 
\begin{equation}
\left\{ 
\begin{array}{c}
\widehat{b}_{k,\alpha }=u_{k,\alpha }\widehat{a}_{k,a}+\upsilon _{k,\alpha }%
\widehat{a}_{-k,a}^{\dagger }, \\ 
\widehat{b}_{k,\alpha }^{+}=u_{k,\alpha }\widehat{a}_{k,a}^{\dagger
}+\upsilon _{k,\alpha }\widehat{a}_{-k,a}.
\end{array}
\right.   \eqnum{A1}
\end{equation}
The Hamiltonian Eq. (\ref{10}) can be written in terms of the quasiboson
operators $\widehat{b}_{k,\alpha }$ and $\widehat{b}_{k,\alpha }^{\dagger }$
as 
\begin{equation}
\widehat{H}=E_c+\widehat{H}_1+\widehat{H}_2,  \eqnum{A2}
\end{equation}
where 
\begin{equation}
E_c=\frac{U_0}{2N_s}N^2+N\left( \varepsilon -zJ\right) +\frac{U_2}{2N_s}%
[\left( N_1-N_{-1}\right) ^2+2N_0(\sqrt{N_1}+\sqrt{N_{-1}})^2]  \eqnum{A3}
\end{equation}
is a constant. $\widehat{H}_1$ and $\widehat{H}_2$ denote the diagonal and
off-diagonal parts, respectively, with

\begin{eqnarray}
\widehat{H}_1 &=&\sum_{k\neq 0}\left( \left[ \left( \overline{\varepsilon }%
_k+\frac{U_0}{N_s}N_0-2\frac{U_2}{N_s}\sqrt{N_{-10}}\sqrt{N_{10}}\right)
(u_{k,0}^2+\upsilon _{k,0}^2)-\left( \frac{2U_0}{N_s}N_0+\frac{4U_2}{N_s}%
\sqrt{N_{-10}}\sqrt{N_{10}}\right) u_{k,0}\upsilon _{k,0}\right] \widehat{b}%
_{k,0}^{\dagger }\widehat{b}_{k,0}\right.   \nonumber \\
&&+\sum_\gamma \left[ \left( \overline{\varepsilon }_k+\frac{U_0}{N_s}%
N_\gamma +\frac{U_2}{N_s}N_\gamma \right) (u_{k,\gamma }^2+\upsilon
_{k,\gamma }^2)-\left( \frac{2U_0}{N_s}+\frac{2U_2}{N_s}\right) N_\gamma
u_{k,\gamma }\upsilon _{k,\gamma }\right] \widehat{b}_{k,\gamma }^{\dagger }%
\widehat{b}_{k,\gamma }+\sum_\gamma \left\{ \left[ \left( -\frac{U_0}{N_s}%
\right. \right. \right.   \nonumber \\
&&\left. \left. +\frac{U_2}{N_s}\right) \sqrt{N_{\gamma 0}}\sqrt{N_{-\gamma
0}}-\frac{U_2}{N_s}N_0\right] (u_{k,\gamma }\upsilon _{k,-\gamma
}+u_{k,-\gamma }\upsilon _{k,\gamma })+\left. \left( \frac{U_0}{N_s}-\frac{%
U_2}{N_s}\right) \sqrt{N_{\gamma 0}}\sqrt{N_{-\gamma 0}}(u_{k,\gamma
}u_{k,-\gamma }+\upsilon _{k,-\gamma }\upsilon _{k,\gamma })\right\} 
\widehat{b}_{k,\gamma }^{\dagger }\widehat{b}_{k,-\gamma }  \nonumber \\
&&+\sum_\gamma \left\{ \left[ \left( \frac{U_0}{N_s}+\frac{U_2}{N_s}\right) 
\sqrt{N_{\gamma 0}}\sqrt{N_{00}}+\frac{2U_2}{N_s}\sqrt{N_{-\gamma 0}}\sqrt{%
N_{00}}\right] \right. (u_{k,\gamma }u_{k,0}+\upsilon _{k,0}\upsilon
_{k,\gamma })-\left( \frac{U_0}{N_s}+\frac{U_2}{N_s}\right) \sqrt{N_{\gamma
0}}\sqrt{N_{00}}(u_{k,\gamma }\upsilon _{k,0}  \nonumber \\
&&\left. \left. +u_{k,0}\upsilon _{k,\gamma })\right\} (\widehat{b}%
_{k,\gamma }^{\dagger }\widehat{b}_{k,0}+\widehat{b}_{k,0}^{\dagger }%
\widehat{b}_{k,\gamma })\right) +const,  \eqnum{A4}
\end{eqnarray}
\begin{eqnarray}
\widehat{H}_2 &=&\sum_{k\neq 0}\left( \left[ \left( \frac{U_0}{2N_s}N_0+%
\frac{U_2}{N_s}\sqrt{N_{-10}}\sqrt{N_{10}}\right) \left( u_{k,0}^2+\upsilon
_{k,0}^2\right) -\left( \overline{\varepsilon }_k+\frac{U_0}{N_s}N_0+2\frac{%
U_2}{N_s}\sqrt{N_{-10}}\sqrt{N_{10}}\right) u_{k,0}\upsilon _{k,0}\right] (%
\widehat{b}_{k,0}^{\dagger }\widehat{b}_{-k,0}^{\dagger }+\widehat{b}_{k,0}%
\widehat{b}_{-k,0})\right.   \nonumber \\
&&+\left\{ \left[ \left( \frac{U_0}{2N_s}-\frac{U_2}{N_s}\right) \sqrt{%
N_{-10}}\sqrt{N_{10}}+\frac{U_2}{N_s}N_0\right] (u_{k,1}u_{k,-1}+\upsilon
_{k,1}\upsilon _{k,-1})-\left( \frac{U_0}{N_s}-\frac{U_2}{N_s}\right) \sqrt{%
N_{-10}}\sqrt{N_{10}}u_{k,1}\upsilon _{k,-1}\right\} (\widehat{b}%
_{k,1}^{\dagger }\widehat{b}_{-k,-1}^{\dagger }+\widehat{b}_{k,1}\widehat{b}%
_{-k,-1})  \nonumber \\
&&+\left[ \frac{U_0}{2N_s}\sqrt{N_{-10}}\sqrt{N_{10}}(u_{k,-1}u_{k,1}+%
\upsilon _{k,-1}\upsilon _{k,1})-\left( \frac{U_0}{N_s}-\frac{U_2}{N_s}%
\right) \sqrt{N_{-10}}\sqrt{N_{10}}u_{k,-1}\upsilon _{k,1}\right] (\widehat{b%
}_{k,-1}^{\dagger }\widehat{b}_{-k,1}^{\dagger }+\widehat{b}_{k,-1}\widehat{b%
}_{-k,1})  \nonumber \\
&&+\sum_\gamma \left\{ \left( \frac{U_0}{2N_s}+\frac{U_2}{2N_s}\right)
N_\gamma \left( u_{k,\gamma }^2+\upsilon _{k,\gamma }^2\right) -\left[ 
\overline{\varepsilon }_k+\left( \frac{U_0}{N_s}+\frac{U_2}{N_s}\right)
N_\gamma \right] u_{k,\gamma }\upsilon _{k,\gamma }\right\} (\widehat{b}%
_{k,\gamma }^{\dagger }\widehat{b}_{-k,\gamma }^{\dagger }+\widehat{b}%
_{k,\gamma }\widehat{b}_{-k,\gamma })  \nonumber \\
&&+\sum_\gamma \left\{ \left( \frac{U_0}{2N_s}+\frac{U_2}{N_s}\right) \sqrt{%
N_{\gamma 0}}\sqrt{N_{00}}(u_{k,\gamma }u_{k,0}+\upsilon _{k,\gamma
}\upsilon _{k,0})-\left[ \left( \frac{U_0}{N_s}+\frac{U_2}{N_s}\right) \sqrt{%
N_{\gamma 0}}\sqrt{N_{00}}+2\frac{U_2}{N_s}\sqrt{N_{-\gamma 0}}\sqrt{N_{00}}%
\right] u_{k,\gamma }\upsilon _{k,0}\right\} (\widehat{b}_{k,\gamma
}^{\dagger }\widehat{b}_{-k,0}^{\dagger }  \nonumber \\
&&+\widehat{b}_{k,\gamma }\widehat{b}_{-k,0})+\sum_\gamma \left\{ \frac{U_0}{%
2N_s}\sqrt{N_{\gamma 0}}\sqrt{N_{00}}(u_{k,0}u_{k,\gamma }+\upsilon
_{k,0}\upsilon _{k,\gamma })\right.   \nonumber \\
&&\left. \left. -\left[ \left( \frac{U_0}{N_s}+\frac{U_2}{N_s}\right) \sqrt{%
N_{\gamma 0}}\sqrt{N_{00}}+2\frac{U_2}{N_s}\sqrt{N_{-\gamma 0}}\sqrt{N_{00}}%
\right] u_{k,0}\upsilon _{k,\gamma }\right\} (\widehat{b}_{k,0}^{\dagger }%
\widehat{b}_{-k,\gamma }^{\dagger }+\widehat{b}_{k,0}\widehat{b}_{-k,\gamma
})\right) .  \eqnum{A5}
\end{eqnarray}
In order to eliminate the off-diagonal part $\widehat{H}_2$ we require that
the coefficients of all terms $\widehat{b}_{k,\alpha }^{\dagger }\widehat{b}%
_{-k,\alpha ^{^{\prime }}}^{\dagger }$ and $\widehat{b}_{k,\alpha }\widehat{b%
}_{-k,\alpha ^{^{\prime }}}$ vanish. In view of condition (12), it is easy
to introduce a set of parameters $\phi _{k,\alpha }$ such that 
\begin{eqnarray}
u_{k,\alpha } &=&\cosh \phi _{k,\alpha },  \nonumber \\
\upsilon _{k,\alpha } &=&\sinh \phi _{k,\alpha }  \eqnum{A6}
\end{eqnarray}
for the convenience of calculation. Conditions (12) and (A6) lead to the
useful relations 
\[
\tanh 2\phi _{k,\alpha }=\frac{2u_{k,\alpha }\upsilon _{k,\alpha }}{%
u_{k,\alpha }^2+\upsilon _{k,\alpha }^2},
\]
\[
\cosh \left( 2\phi _{k,\alpha }\right) =u_{k,\alpha }^2+\upsilon _{k,\alpha
}^2=\frac 1{\sqrt{1-\tanh ^22\phi _{k,\alpha }}},
\]
with which the Hamiltonian (10) can be finally reduced to the diagonal form
as given in Eq. (\ref{13}).

\end{document}